\documentclass[lettersize,journal]{IEEEtran}
\usepackage{amsmath,amsfonts}
\usepackage{algorithmic}
\usepackage{algorithm}
\usepackage{array}
\usepackage[caption=false,font=normalsize,labelfont=sf,textfont=sf]{subfig}
\usepackage{textcomp}
\usepackage{stfloats}
\usepackage{url}
\usepackage{verbatim}
\usepackage{graphicx}
\usepackage{cite}
\begin{document}

\title{A PDD-Inspired Channel Estimation Scheme in
NOMA Network}

\author{Sumita Majhi, and Pinaki Mitra 
\thanks{Sumita Majhi and Pinaki Mitra are with Department of Computer Science and Engineering
Indian Institute of Technology Guwahati,
Guwahati, Assam, India
\{sumit176101013, pinaki\}@iitg.ac.in}

\thanks{Manuscript received April 19, 2021; revised August 16, 2021.}}

\markboth{Journal of \LaTeX\ Class Files,~Vol.~14, No.~8, August~2021}%
{Shell \MakeLowercase{\textit{et al.}}: A Sample Article Using IEEEtran.cls for IEEE Journals}


\maketitle

\begin{abstract}
In 5G networks, non-orthogonal multiple access (NOMA) provides a number of benefits by providing uneven power distribution to multiple users at once. On the other hand, effective power allocation, successful successive interference cancellation (SIC), and user fairness all depend on precise channel state information (CSI). Because of dynamic channels, imperfect models, and feedback overhead, CSI prediction in NOMA is difficult. Our aim is to propose a CSI prediction technique based on an ML model that accounts for partially decoded data (PDD), a byproduct of the SIC process. Our proposed technique has been shown to be efficient in handover failure (HOF) prediction and reducing pilot overhead, which is particularly important in 5G. We have shown how machine learning (ML) models may be used to forecast CSI in NOMA handover 
\end{abstract}



\maketitle

\section{INTRODUCTION}
\IEEEPARstart{N}{on-orthogonal} multiple access \cite{20}, an information theoretic approach incredibly useful for 5G networks, serves numerous users at once by using uneven power allocation. To provide strong signals for the user with a poor channel and lesser signals for the user having a strong channel, accurate CSI is crucial in determining the ideal power levels for each user. Power allocation may be ineffectual without precise CSI, which would result in poor signal reception for both users. When using NOMA, a user decodes its own signal first, cancels out the stronger user's signal interference, and then decodes the weaker user's signal. For interference cancellation to be successful, accurate CSI is essential. Relative interference caused by inaccurate CSI can severely impair the weaker user's performance. Ensuring user fairness is the goal of NOMA. The system's ability to allocate resources efficiently depends on accurate CSI, which guarantees that both strong and weak users receive a sufficient level of signal quality. Due to variables including user movement, obstructions, and fading, wireless channels are naturally dynamic, meaning that signal strength and propagation characteristics can change quickly. Accordingly, it is challenging to make very accurate predictions for future CSI. The intricacies of real-world situations might not be fully captured by conventional channel models \cite{21}. To further complicate the modeling procedure, NOMA incorporates non-linear behavior because of power allocation and SIC. Getting regular input from every user on the CSI might add a lot of load to the network. It is important to strike a compromise between minimizing feedback overhead and making accurate predictions.
\par
Handover failures \cite{11}, \cite{18} in NOMA networks happen when a user switches between cell towers and needs to pair with a new device in the next cell base station, as shown in Fig. 1, but the handoff doesn't go through. These can result in dropped calls, slow data connections, and an unpleasant user experience. Predicting CSI is essential to ensuring seamless handovers. The network is able to make educated judgments about when and to which cell tower to hand off a user by precisely forecasting future channel conditions. By doing this, the likelihood of a faulty handover and the feared "ping-pong" effect in which a user switches between two towers as a result of poor handover decisions is decreased. Essentially, smooth handovers and enhanced network performance are dependent on accurate CSI prediction.
\par
Assume that we are in a NOMA downlink network and that we are receiving a signal that contains data from two users (User A and User B). The signal from User A is stronger than the signal from User B. We make an effort to decode User A's signal during the decoding phase fully. SIC is not currently involved in this procedure. Next, we run SIC using the calculated CSI. We attempt to "subtract" User A's strong signal effect from the total signal that was received. Ideally, this leaves us with the message from User B alone. However, the CSI estimations in the downlink NOMA network could not be accurate owing to constraints in real-world circumstances. It is possible for the channel to be dynamic and to evolve over time. Therefore, User A's involvement may not be entirely eliminated by the SIC procedure. The residual interference and the partially recovered user B's information can constitute partially decoded data, which we refer to for the rest of the paper as PDD. PDD provides useful hints about the current channel status, even if it does not immediately reveal the entire CSI. We may evaluate the success of the SIC process by looking at the amount of residual interference found in PDD. By detecting channel dynamics, the knowledge obtained from evaluating PDD may be applied to forecast future channel behavior. Excessive residual interference may be a sign of abrupt channel shifts.  Data that has been partially decoded offers a constant feedback loop on the channel state. We can observe changes in the channel's behavior over time and predict future modifications by observing residual interference levels. We can increase SIC performance in succeeding transmissions by achieving more effective SIC by refining future predictions and recognizing the limitations of existing CSI estimates through residual interference analysis. When it comes to Adaptive Communication, the real-time feedback loop enables dynamic modifications to SIC methods and power allocation, guaranteeing effective communication even under shifting channel circumstances. Additionally, it lessens the feedback overhead needed to comprehend the channel without necessitating complete decoding for each transmission, which may lessen network overhead and eventually decrease the usage of pilot overhead over time.
\par
All prior studies, including \cite{1}, \cite{3}, \cite{4}, and \cite{5}, have focused on the linear minimum mean square error (LMMSE) or Minimum Mean Square Error (MMSE) approach. However, these methods are computationally complicated and need a costly matrix transposition methodology. When dealing with devices that have limited resources, it is important to strike a trade-off between accuracy and resource constraints. Both the LMMSE and MMSE approaches need to be able to understand the noise variation in the channel. However, in real-world situations, we often encounter dynamic noise levels that are challenging to measure accurately. Inaccuracies in noise estimation might result in less than ideal performance of certain channel estimation algorithms.  Another limitation of LMMSE/MMSE systems is that they presume the channel coefficients stay generally constant over the estimation time. In situations characterized by rapid fluctuations in the channel (fast fading), the channel might undergo substantial changes before the estimate is used, hence diminishing its efficacy. In contrast, machine learning solutions have advantages such as flexibility, adaptability, and decreased computing load, making them more suitable for real-world applications. It has the capacity to overcome these restrictions via the provision of adaptive, data-driven, and computationally efficient solutions for channel estimation. To the best of our knowledge, no work has been done in NOMA-HO taking PDD as CSI with the help of the ML model to predict future channel estimation. 

\subsection{ABBREVIATIONS}
\begin{table}[ht]
\caption{Abbreviation Table}
\begin{center}
\begin{tabular}{|c|c|}
\hline
 Abbreviation &  Definition \\
 \hline
 NOMA  & Non-Orthogonal Multiple Access \\ 
 \hline
 SIC  & Successive Interference Cancellation \\ 
 \hline
  CSI  & Channel State Information \\ 
 \hline
  PDD  & Partially Decoded Data  \\ 
 \hline
  HOF  & Handover Failure \\ 
  \hline
 ML  & Machine Learning \\ 
 \hline
  RSRP  &  Reference Signal Receive Power \\ 
 \hline
  RSRQ  & Reference Signal Receive Quality \\ 
 \hline
  SNR  & Signal-to-Noise Ratio \\ 
 \hline
  CQI  & Channel Quality Indicator \\ 
 \hline
PMI & Precoding Matrix Indicator  \\
  \hline
  RI & Rank Indicator  \\ 
 \hline
 LMMSE  & Linear Minimum Mean Square Error   \\ 
 \hline
  MMSE  & Minimum Mean Square Error \\ 
  \hline
 HOF & Handover Failure  \\ 
 \hline
  RLF & Radio Link Failures \\ 
 \hline
  SINR  & Signal-to-Interference-plus-Noise Ratio  \\ 
 \hline
 NOMA-HO  & NOMA Handover  \\ 
 \hline
 UE  & User Equipment  \\ 
 \hline
DL & Deep Learning \\
\hline
SC & Superposition Coding \\
\hline
 BS & Base Station \\
\hline
CNN & Convolutional Neural Network \\
 \hline
 RNN-LSTM & Recurrent Neural Network-Long Short-Term Memory \\ 
 \hline
  SINR  & Signal-to-Interference-plus-Noise Ratio  \\ 
 \hline
 NRMSE  & Normalized Root Mean Square Error   \\ 
 \hline
 RMSE  & Root Mean Square Error  \\ 
 \hline
MASE & Mean Absolute Scaled Error  \\
\hline
MSE & Mean Squared Error \\
\hline
 $R^{2}$ & R-squared  \\
  \hline
  MAE & Mean Absolute Error \\
  \hline
 TTT & Time to Trigger \\
  \hline
 BER & Bit Error Rate \\
   \hline
\end{tabular}
\end{center} 
\end{table}

\subsection{Motivation and contributions}
The ever-increasing demand for reliable and high-speed wireless connectivity in NOMA networks presents a significant challenge: accurately predicting channel behavior.  Traditional methods struggle to account for the dynamic nature of radio channels, leading to potential service disruptions due to handover failures (HOF) and radio link failures (RLF).  This research explores the potential of utilizing a rich set of channel metrics, including Reference Signal Received Quality (RSRQ), Signal-to-Interference-plus-Noise Ratio (SINR), PDD, and Channel Quality Indicator (CQI), to develop a more robust approach to channel prediction. 
\par
To the best of our knowledge, no work has been done on taking channel state information to predict channel status from the perspective of NOMA handover (NOMA-HO) on the cellular network. The objective of this study is to examine methodologies that optimize computing requirements while maintaining precision. The circumstances of a channel are naturally subject to change over time. Our motivation stems from the need to develop adaptive channel estimation algorithms that possess the ability to continuously learn and adapt in the face of dynamic settings, thereby guaranteeing resilient performance. Also, predicting the ideal channel state information is crucial for achieving optimum performance in NOMA networks. This information may effectively address the issues of HOF and RLF. The previous research included many forms of CSI, such as RSRP, RSRQ, SNR, CQI, precoding matrix indicator (PMI), and rank indicator (RI) \cite {6}\cite {7}\cite{8}. However, none of them considered PDD as an additional kind of CSI for predicting channel status. PDD is a strong, unusual CSI measure. Instead of dedicated signaling, it estimates channels using user traffic. PDD shows channel performance more realistically than specialized signaling systems since it matches actual user traffic. This improves channel prediction, particularly for different traffic types. It can also easily adapt to changes in user traffic. The summary of our contributions is as follows:
\begin{itemize}
    \item Developed an ML model to predict CSI accounting PDD.
    \item Shown the application of the CSI prediction model towards minimizing handover failure (HOF) rate, ping-pong rate, and number of false alarms. Conduct system-level simulations to analyze how inaccurate channel information due to limited CSI contributes to poor network performance.
   \item Shown the application of the CSI prediction model to reduce the need for specialized pilot signals. system-level simulations show the superiority of our work in reducing the MSE and BER performance of SNR.
   \item Given the limited size of the dataset, we have created artificial data. Simulation results exhibit the competitive performance of the proposed model in comparison to other ML models.

\end{itemize}
\section{Related work}
NOMA is a proven technology in cellular networks. However, research on NOMA handover processes is sparse. Due to the complexity of handling several users with different power levels in a handover situation with various antennas, this sector has seen little optimization. Further study into NOMA network handover techniques is possible due to this gap in the literature. Prior studies \cite{8} \cite {9}\cite {10}\cite {11}\cite {12} \cite{13} on channel estimation in NOMA networks mostly depend on CSI-based channel estimation. Nevertheless, this method has notable disadvantages. The approach, \cite{13} discussed in the study, deals with user mobility and handover in cell-free massive MIMO networks. However, its main emphasis is on minimizing pilot changes and computing complexity to mitigate the pilot contamination problem. It uses the same pilot sequence by multiple UEs, which can reduce the accuracy of channel estimation and further reduce the handover prediction accuracy. The study \cite{9} focuses on analyzing the energy efficiency of hybrid VLC-RF systems that use NOMA. However, it does not directly address the topic of handover in NOMA networks. Although the approach takes into account user mobility and inadequate channel information, it does not address the difficulties of inaccurate channel prediction that are experienced during handover. This paper \cite{10} investigates the use of NOMA for soft handover in multi-cell networks, with the specific goal of enhancing the data rates for users near the cell edges. The purpose of this is to set the necessary circumstances for a successful handover while also ensuring that the overall network throughput is maintained. However, the approach fails to include CSI prediction, which is an important parameter that negatively impacts the accuracy of handover prediction. The study in \cite{11} investigates the implementation of NOMA for handover in cellular networks, with a specific emphasis on enhancing the throughput of users at the cell edge. The theoretical criteria for a successful handover have been met. However, the paper does not specifically mention or work on predicting channel information. "Soft handover"—where a user connects to the serving and target base stations simultaneously during handover, has been mentioned in the work as necessary. It necessitates quick and precise channel estimation in order for both base stations to assess user data, and signal strength has been missing. The article in \cite{12}  focuses on the difficulties associated with channel estimation for NOMA with beamforming in high-mobility Vehicle Communication Networks (VCN) and highlights the problems due to high mobility. It suggests an approach to deal with the problem of outage probability limitations that makes use of semidefinite relaxation (SDR) and convex restrictions. Although the research admits inaccurate CSI, it doesn't explore how to make the channel estimate procedure better. Its main objective is to address the shortcomings of the current estimations. 
\par
Traditional CSI parameters like RSSI, RSRQ, CQI, PMI, and SINR provide valuable channel information, but they often fail to predict the dynamic behavior of wireless channels. RSSI is affected by factors other than channel quality, RSRQ only shows signal reception quality, CQI only shows potential data transmission rates, PMI only estimates MIMO system precoding, and SINR only measures noise and interference. The paper in \cite{8} presents an innovative and possibly influential method for detecting human motion and recognizing activities by using RSSI, which quantifies the intensity of the WiFi signal received in smart homes. CSI provides a plethora of information that extends beyond signal strength. Although RSSI is influenced by human motion, it may also be impacted by variables such as proximity to the router, reflections off walls, and network congestion.
\par
A Deep Learning (DL) based approach for channel state estimation in a heterogeneous network using Long Term Evolution (LTE) Radio Interfaces is described in \cite{2}. The suggested approach uses historical data to train the deep learning model. The necessity of striking a balance between historical data and model complexity is mentioned in the paper, which might restrict their use in situations when there is a lack of data. The use of pre-measured radio channel metrics, such as SINR, SNR, and RSSI, to forecast transmission quality is mentioned in the text. Although this is helpful, it may not adequately convey the dynamic nature of the channel, particularly in networks with different topologies. The paper in \cite{6} recognizes that complex-valued CSI data cannot be directly processed using CNN techniques that were created for real-valued data (such as image processing). Their approach, which might be ineffective, is to divide the actual and imaginary components into distinct columns. The accuracy of categorization may be impacted when splitting complex numbers since the natural links between the real and imaginary components are lost. The lack of real-time channel information in the study can help reduce the difficulty of the suggested approach. It's possible that Deep Learning models in \cite{7} developed for particular channel circumstances won't adapt well to new situations. When applied to various real-world contexts, this may result in decreased accuracy. 
Table II provides a comprehensive comparison between the prior research and the research being presented. 

\begin{table*}[ht]
\caption{Comparison Table}
\begin{center}
\begin{tabular}{|c|c|c|c|c|}
\hline
 Reference &  LMMSE /MMSE & Consider PDD as CSI  & Deep Learning Model & Consider CSI in NOMA-HO  \\
 \hline
 \cite{1}  & $\checkmark$ & $\checkmark$ & $\times$ & $\times$\\ 
 \hline
 \cite{2}  & $\times$ & $\times$  & $\checkmark$ & $\times$\\ 
 \hline
 \cite{3}  & $\checkmark$ & $\checkmark$  & $\times$ & $\times$\\ 
 \hline
  \cite{4}  & $\checkmark$ & $\checkmark$  & $\times$ & $\times$\\ 
 \hline
  \cite{5}  & $\checkmark$ & $\checkmark$  & $\times$ & $\times$\\ 
  \hline
 \cite{6}  & $\times$ & $\times$  & $\checkmark$ & $\times$\\ 
 \hline
  \cite{7}  & $\times$ & $\times$  & $\checkmark$ & $\times$\\ 
 \hline
  \cite{8}  & $\times$ & $\times$  & $\times$ & $\times$\\ 
 \hline
  \cite{10}  & $\times$ & $\times$  & $\times$ & $\times$\\ 
 \hline
  \cite{11}  & $\times$ & $\times$  & $\times$ & $\times$\\ 
 \hline
 Proposed work & $\times$ & $\checkmark$  & $\checkmark$ & $\checkmark$\\
  \hline
\end{tabular}
\end{center} 
\end{table*}

\section{System model}
Within the scope of a downlink NOMA system, the analytical model consists of a base station of $M$ antennas and $N$ user equipment (UEs), each equipped with a single antenna. The Rayleigh fading channel model is used to describe the probabilistic behavior of wireless channels. The formulation of the received signal $y$ is achieved by taking into account parameters such as channel matrices $h_{i} = \left\{h_{1},h_{2},\ldots, h_{N}\right\}$, transmitted signals $x_{i} =  \left\{x_{1},x_{2},\ldots, x_{N}\right\} $, $ \forall i \in \left\{1, 2, \ldots, N \right\}$, and the additive white Gaussian noise $n$. The equation is denoted as:
\begin{equation}
 y = \sum_{i=1}^{N} h_{i}x_{i} +  n 
\end{equation}
We can rewrite Equation 1:
\begin{equation}
  y_{1} = \underbrace{h_{1}x_{1}(j)}_{User 1} + \sum_{i\neq1}^{N} h_{i} x_{i}(j) + n_{1}(j)
\end{equation}
where $j$ indicates the time index and $j \in \left\{1,2, \ldots, J \right\}$. 
\par
In this article, we have taken a new parameter, PDD, as a supplementary CSI metric. The analytical equation is expressed below. To understand this, we have taken two user scenarios in a NOMA network for easy understanding; user 1 indicated near user, and user 2 indicated far user from BS. The equation for superposition coding (SC) at the transmitter (BS):
\begin{equation}
    y' = h_{1}x_{1}(j) + h_{2}x_{2}(j) + n_{1}(j)
\end{equation}

The equation at user 2 is denoted as:
\begin{equation}
    y' = h_{1}x_{1}(j) + h_{2}x_{2}(j) + \hat{x_{2}}(j-1) + n_{1}(j)
\end{equation}
At user 2, it decodes message $x_{2}$ by treating user 1's message as noise. $\hat{x_{2}}(j-1)$ indicates PDD information in the previous time step.
\par
At user 1, it decodes user 2's message first. subtracts it, then decodes its own message. The equation at user 1 is denoted as:
\begin{equation}
    y' = h_{1}x_{1}(j) + h_{2}x_{2}(j) + \hat{x_{1}}(j-1) + n_{1}(j)
\end{equation}
where $\hat{x_{1}}(j-1)$ indicates PDD information in the previous time step. Within the context of NOMA networks operating in the Rayleigh fading channel model, the need for efficient channel estimation to make the handover choice becomes apparent. As the number of UEs increases, the likelihood of user pairing based on handover decisions to improve the total sum rate also becomes difficult. In order to guarantee the resilience and effectiveness of these networks, it is important to cultivate channel status and different CSI parameters in the NOMA network. Fig. 1 illustrates the concept of the NOMA handover. We have considered a two-cell scenario for ease of understanding. We have checked the condition for making a handover decision, $CSI_{0} + PDD_{0} < CSI_{1} + PDD_{1}$, then the handover has been made, and user $U_{0}^{F}$ will connect to the base station $BS_{1}$.
\begin{figure}[htbp]
    \centering
    \includegraphics[width=0.5\textwidth]{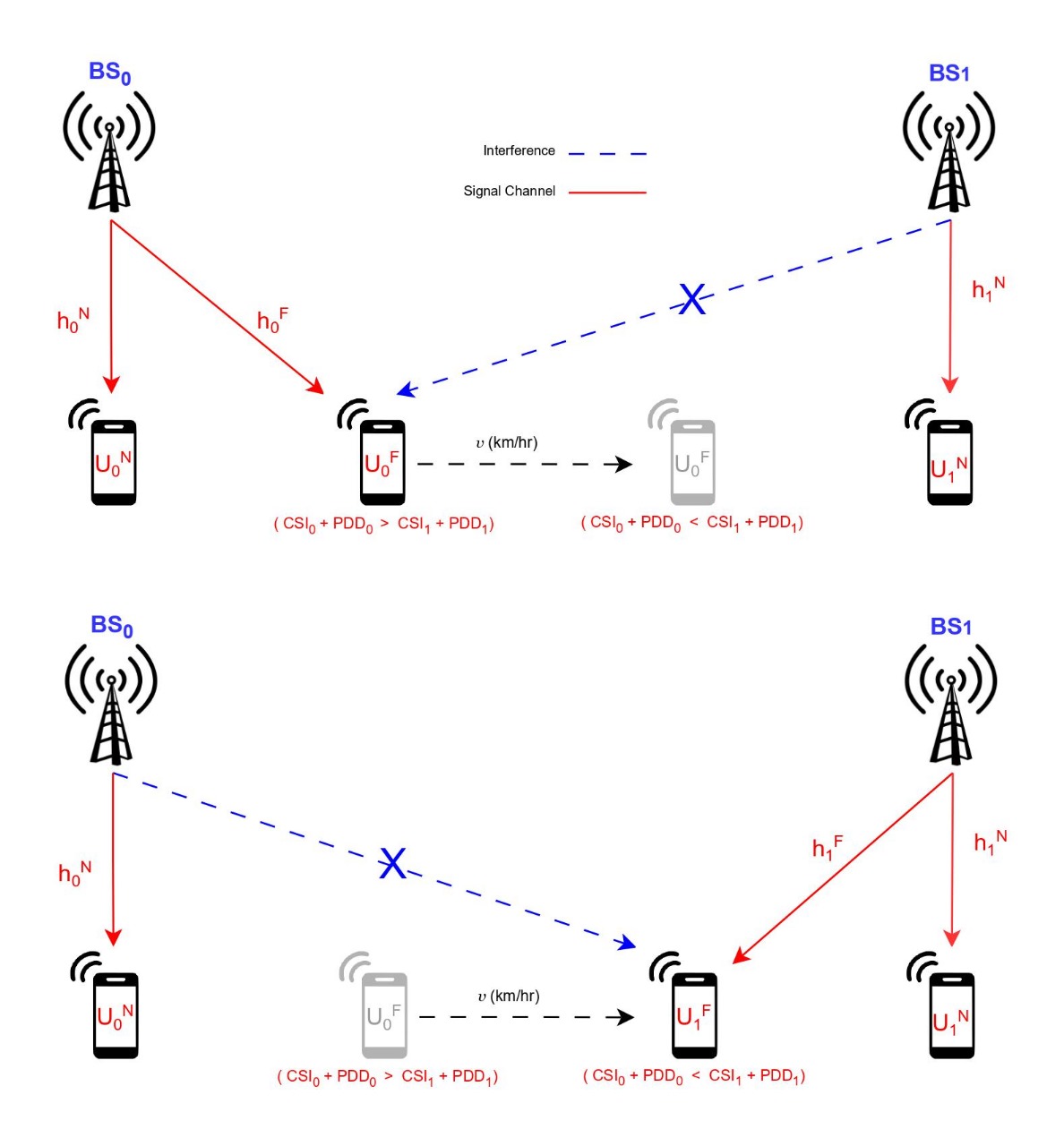} 
    \caption{System model of NOMA-HO}
    \hfill
\end{figure}
Fig. 2 illustrates the architecture of the NOMA transceiver using the RNN-LSTM model. The exclusive error information for channel status will be given after the demodulation state to the SIC/PIC module through the RNN-LSTM model to predict channel status. Channel prediction relies on exploiting information extracted after demodulation but before final data decoding. Following CP removal and FFT, demodulation recovers the symbols, potentially containing channel errors. These errors themselves become valuable clues. By analyzing these post-demodulation symbols with error patterns, techniques can estimate the channel's characteristics (fading, noise) and create CSI to predict future channel behavior and improve transmission reliability.
\section{Proposed RNN-LSTM model}
\begin{figure}[htbp]
    \centering    \includegraphics[width=0.9\linewidth, height=0.7\linewidth]{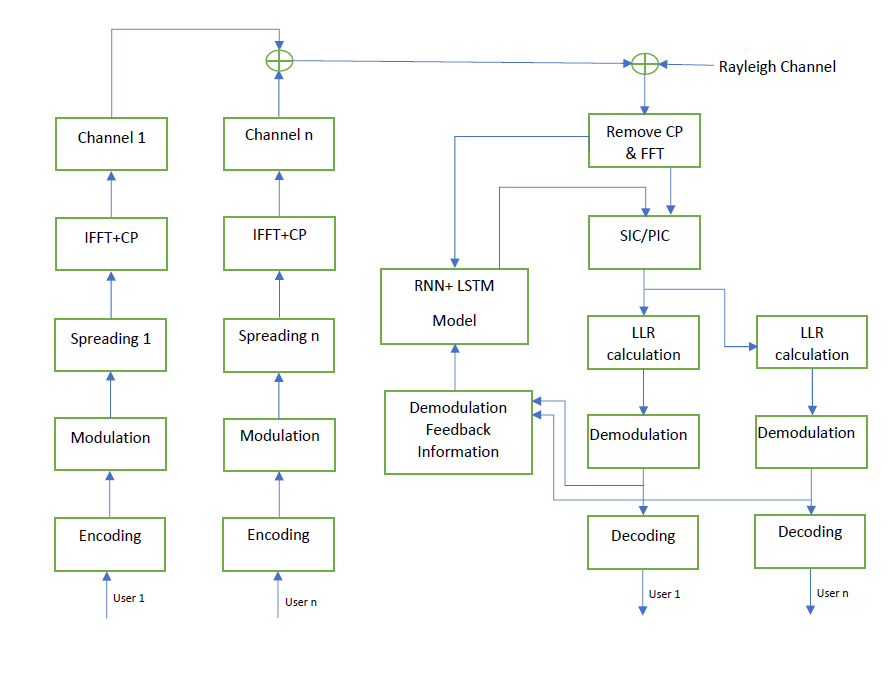} 
    \caption{System model of NOMA transceiver with RNN-LSTM model}
    \label{fig:noma_transceiver}
\end{figure}
\subsection{Data Preperation}
Algorithm 1 outlines a transfer learning approach for predicting future channel state information in a wireless communication system using RNNs. It leverages a pre-trained RNN model, originally trained on a different but relevant task, to extract informative features from current and historical channel measurements CSI data like RSRQ, CQI, SNR, and PDD). These features are then fed into a newly trained regressor model to predict future values. This approach aims to improve channel prediction accuracy by utilizing the knowledge learned by the pre-trained RNN model and fine-tuning it for the specific task of channel state prediction. Fig 3 provides a pictorial representation of the workflow outlined in Algorithm 1. As evidenced in Fig 4, 5, and 6 (data source: \cite{15}), we can visualize the impact of data augmentation. Fig 4 presents the initial distribution of the data, while Fig 5 showcases the transformed distribution after augmentation. To enhance our understanding of these changes, Fig 6 utilizes kernel density estimation (KDE) plots, providing a smoother representation of the data distributions in both figures. This comparative analysis allows for a clear assessment of how augmentation has modified the data. The new data set will be available in \cite{16}. The number of epochs $E$ and batch size $B$ determine the number of times the regressor model processes the features extracted from the pre-trained model $h$. The total complexity depends on the product of these factors multiplied by the complexity per batch $T_{new}(h)$. The total time complexity of the algorithm becomes $O(E * B * T_{new}(h))$.

\begin{algorithm}
\caption{Transfer Learning Algorithm to generate Regressor model.}\label{alg:rnn_channel_prediction}
\begin{algorithmic}
\STATE {\textsc{Input:}}
\STATE \hspace{0.5cm}$ \textbf{Pre-trained RNN model}$ $\mathbf{M}_{pre}(\mathbf{x}; \mathbf{\theta}_{pre})$
\STATE \hspace{0.5cm}$ \textbf{Training data set}$ $\mathbf{x}_i$ 
\STATE \hspace{0.5cm}$i \in  \left\{RSRQ, CQI, PDD, SNR\right\}$ 
\STATE {\textsc{Output:}}
\STATE \hspace{0.5cm}$ \textbf{Regressor model}$ $\mathbf{M}_{new}(\mathbf{h}; \mathbf{\theta}_{new})$ 
\STATE {\textsc{Feature Extraction:}} $M_{pre}(x; \theta_{pre})$ takes an input vector $x$ containing current and historical CSI measurements and outputs $h$.
\end{algorithmic}

\[
h = M_{pre}(x; \theta_{pre})
\]

\begin{algorithmic}
\STATE {\textsc{Regressor Model:}} $M_{new}(h; \theta_{new})$ takes the extracted hidden state sequence $h$ as input and predicts $\hat{y}$, where $y_{i} \in  \left\{  RSRQ, SNR \right\}$.
\end{algorithmic}

\[
\hat{y} = M_{new}(h; \theta_{new})
\]

\begin{algorithmic}
\STATE {\textsc{Training the Regressor model:}} $M_{new}(h; \theta_{new})$ is trained by minimizing the MSE loss function $L$.
\end{algorithmic}

\[
L(\hat{y}, y) = (y - \hat{y})^2
\]

\textbf{Note:} Here, $y$ represents the actual RSRQ and SNR values.
\label{alg:rnn_channel_prediction}
\end{algorithm}

\begin{figure}[htbp]
    \centering
    \includegraphics[width=0.5\textwidth]{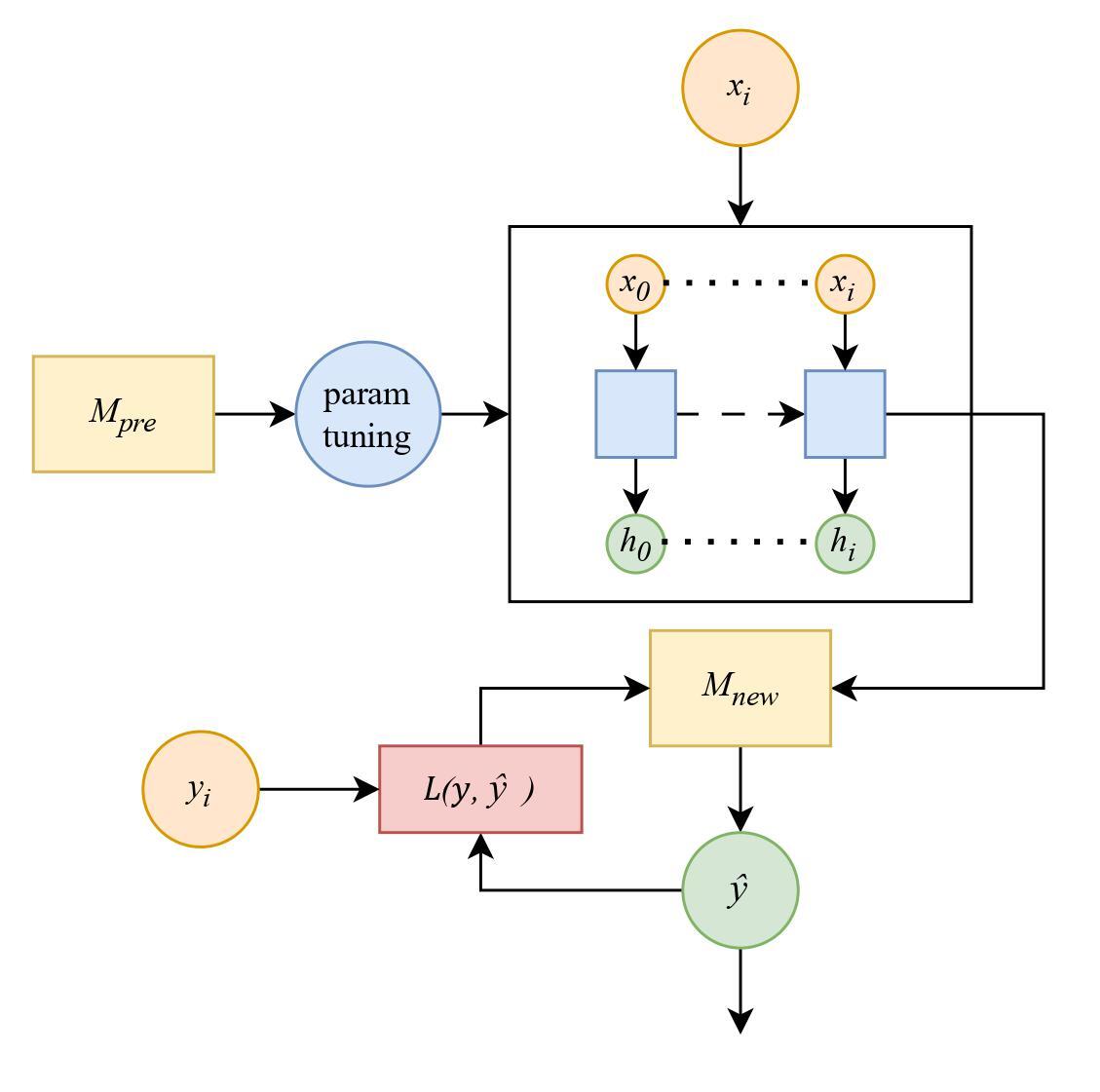} 
    \caption{Flowchart of Algorithm 1}
    \hfill
\end{figure}
\begin{figure}[htbp]
    \centering
    \includegraphics[width=0.5\textwidth]{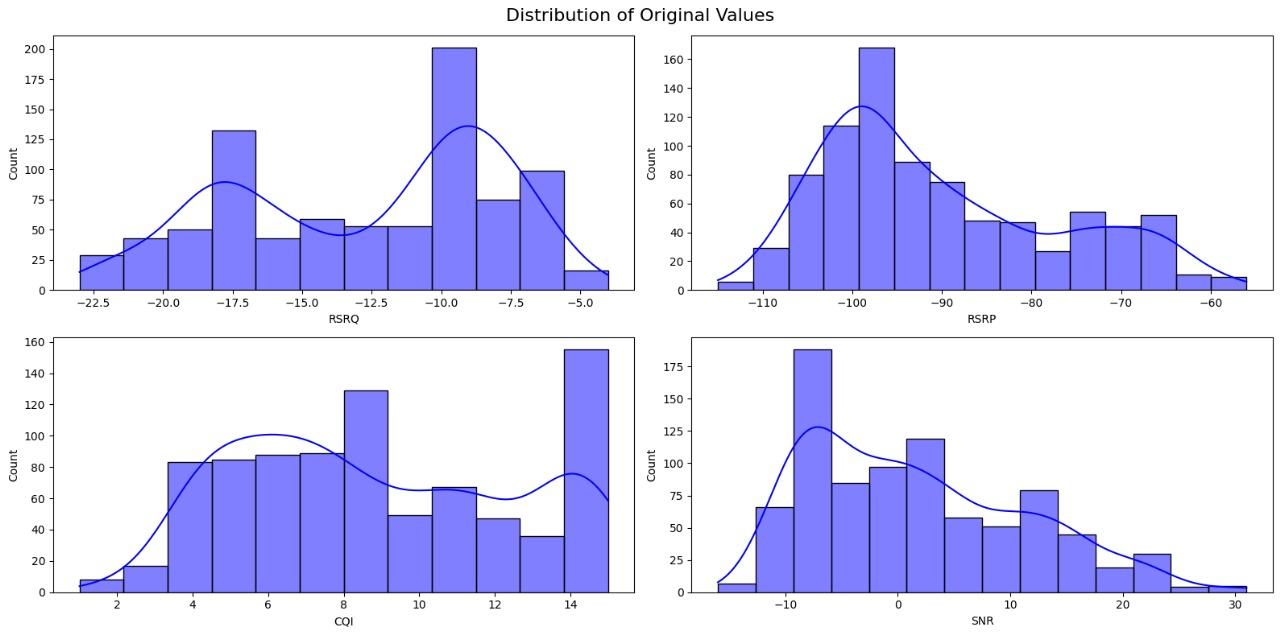} 
    \caption{Distribution of Original Values}
    \hfill
\end{figure}
\begin{figure}[htbp]
    \centering
    \includegraphics[width=0.5\textwidth]{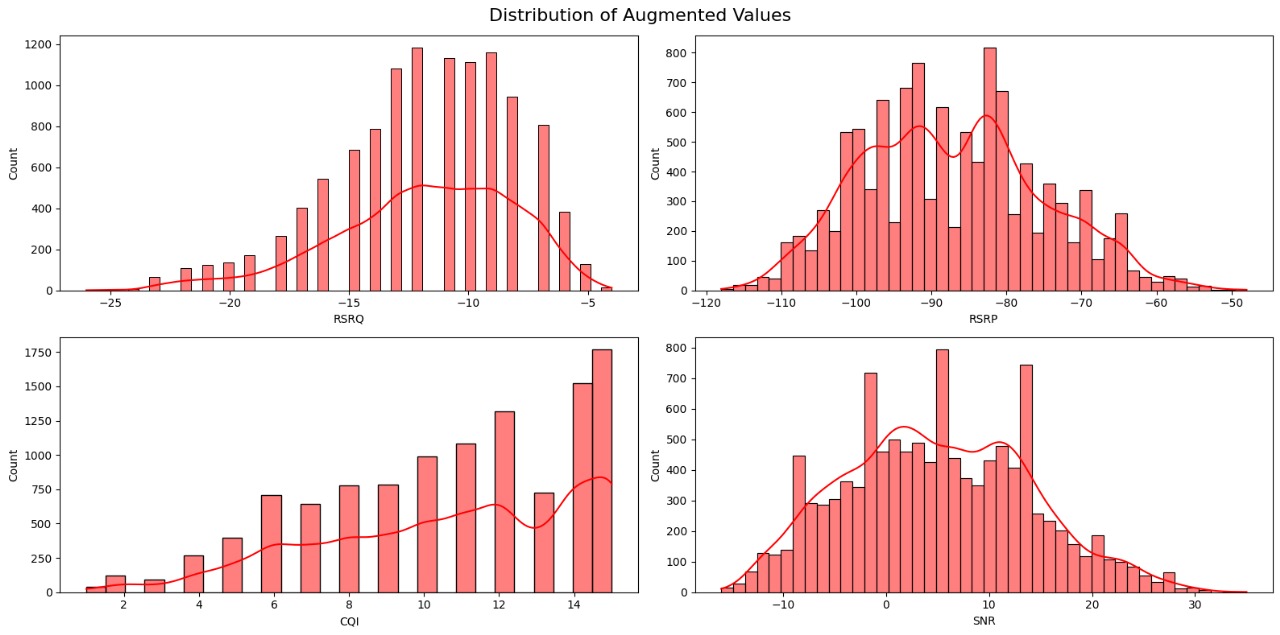} 
    \caption{Distribution of Augmented Values}
    \hfill
\end{figure}
\begin{figure}[htbp]
    \centering
    \includegraphics[width=0.5\textwidth]{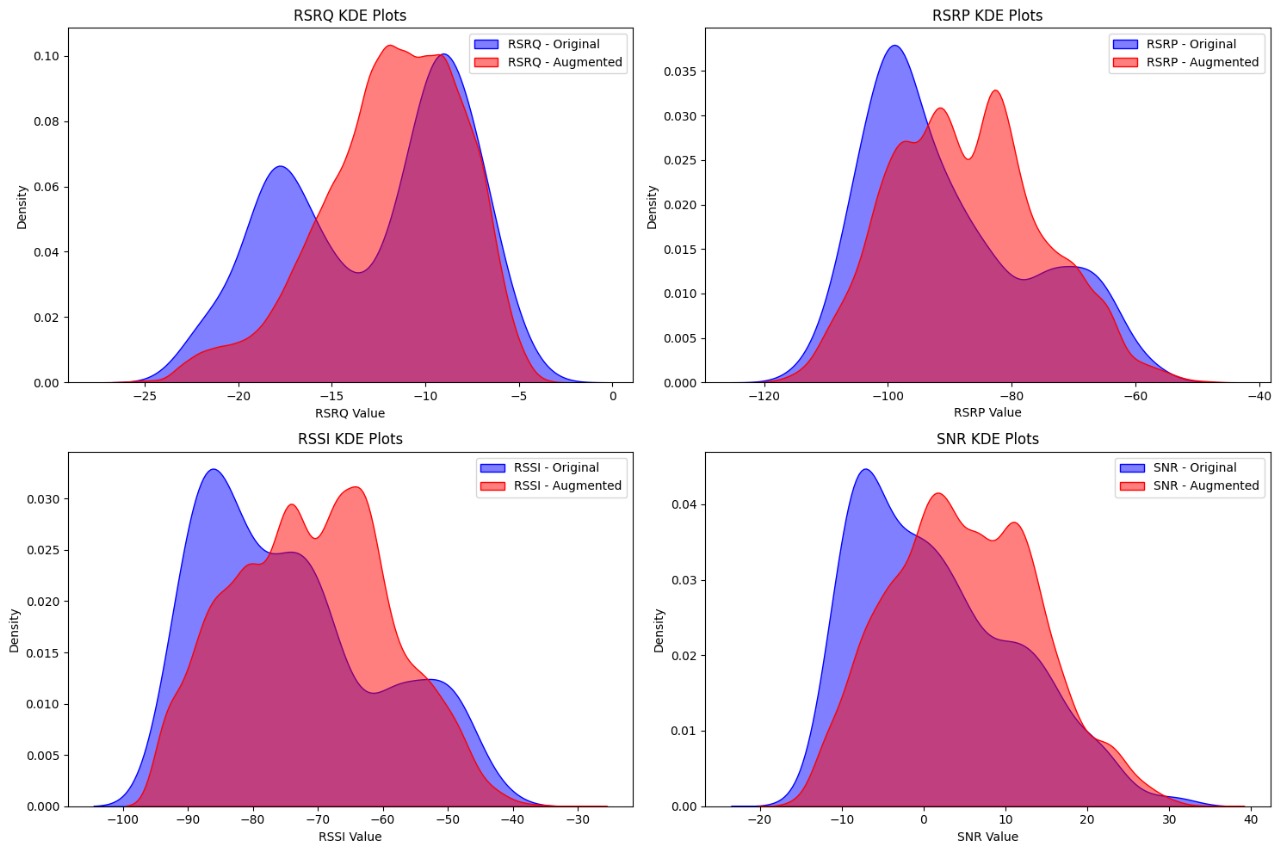} 
    \caption{KDE Plots}
    \hfill
\end{figure}
\subsection{Complexity Analysis}
The complexity of a simple RNN-LSTM model is determined by examining the comparable operations that are involved. The fundamental operations of RNN-LSTMs depend on performing matrix multiplications and activation functions inside a backpropagation algorithm during the training process. An RNN-LSTM architecture comprises an RNN layer, often including LSTMs, followed by a final dense layer to provide the output. The internal gate actions (input, forget, output) and cell state update inside each LSTM unit contribute to the complexity every time step. The commonly used notation for this complexity is $O(D_i * D_o + D_o^2)$, where $D_i$ represents the input dimension (the number of features in the input vector) and $D_o$ represents the output dimension (the number of hidden units in the LSTM). The complexity of the final dense layer stays constant at $O(D_o * D_{out})$, where $D_o$ represents the output dimension from the RNN layer (number of hidden units) and $D_{out}$ represents the number of neurons in the output layer (typically 1 for RSRQ/SNR prediction). RNN-LSTMs sequentially process data. Hence, the intricacy increases with each successive time step T, whereas S is the number of training samples (iterations). The RNN-LSTM model's overall computational complexity is estimated as follows:

 $O(T * S * (D_i * D_o + D_o^{2} + D_o * D_{out}))$

 The primary factors contributing to its complexity are the quantity of time steps $(T)$ and the calculations performed inside the LSTM unit $(D_i, D_o)$. The level of complexity may exhibit variability based on the number of hidden units $(D_o)$ and the input sequence length $(T)$.
 
 \section{Results and Discussion}
This research investigates three key aspects of NOMA networks. The first part explores the application of an RNN-LSTM model for improving CSI prediction. The second part analyzes the relationship between handover frequency (switching between cell towers) and UE speeds within the network. The third part compares BER performance with existing pilot-based channel estimation techniques. All of the given pronged approaches aim to optimize network performance in NOMA systems by understanding handover behavior at different UE speeds and leveraging machine learning for more accurate CSI prediction.

\subsection{Simulation Setup and Results}
This study presents a feasible parameter configuration for the NOMA system model in the context of 5G. In this case, we have two BSs that are responsible for servicing two different cells. The number of Users per Cell is 2. Each Base Station has 4x4 antenna configurations.  Each UE is equipped with a single antenna. In Table III, we have mentioned a precise description of network parameter configuration. We will establish precise channel models for the base station-to-user equipment connections, taking into account issues like route loss and fading. Regarding the fading feature, we will make the assumption that the distribution of $x(t)$ conforms to a Rayleigh distribution, which is often used to model real-world fading situations. The simulation will assess the NOMA system's performance under several scenarios, such as altering the Transmit SNR at the BSs and the power allocation factor of different UEs within each cell. The precise values for these parameters will be specified in Table III. 
\par
The number of hidden units (Ncell) in the LSTMs is set at 16. In addition, configure distinct activation functions such as Rectified Linear Unit (ReLU) for the Long Short-Term Memory (LSTM) units. Using three LSTM layers might be advantageous for capturing intricate temporal connections. The duration of the input sequence, referred to as the look-back window, also has a significant impact. Usually, the window size is set at L = 10 in order to provide context. Ultimately, using methods such as dropout regularization with a dropout rate of around 0.2 may effectively mitigate overfitting by randomly eliminating neurons throughout the training process.
\begin{table}
\caption{ Parameters setup}
\begin{center}
\begin{tabular}{||p{2cm}|p{4cm}|p{1.5cm}||}

\hline
 Parameters &  Names & Values    \\[0.5ex] 
 \hline
 $K$  & Number of time steps& 1000  \\ 
 \hline
$t_{i}$ & Number of input time steps & 10 \\
\hline
$t_{o}$ & Number of future time steps &  1\\
\hline
 N  &  Number of users & 4  \\ 
 \hline
$ B$  & Number of BS & 2  \\ 
 \hline
  $C$  &  Number of cells & 2  \\ 
 \hline
  $v$  & Uniform speed per one time step & 0.1m  \\ 
  \hline
  $\gamma$ & Transmit SNR &  -9dB to 14dB\\
  \hline
   $\rho $ & Transmit RSRQ & -8dB to -20dB \\
  \hline
   $d_{bn}$  & Distance from BS to near user & 20  \\ 
 \hline
  $d_{bf}$  & Distance from BS to far user & 50  \\ 
 \hline
 $ M$  &Number of antennas at BS & 4  \\ 
 \hline
 $a$ & Number of antennas at UE& 1  \\
  \hline
 
\end{tabular}
\end{center}

\end{table}

\subsection{Performance Metrics}
A mix of scale-free and scale-dependent measurements will be the most appropriate performance metrics. Normalized Root Mean Square Error (NRMSE) is a kind of scale-free metric. The Root Mean Square Error (RMSE) is normalized by partitioning it by the standard deviation of the actual target values $(RSRQ/SNR)$. This facilitates the comparison of performance by using datasets that include varying scales. Standardized RMSE and Mean Absolute Scaled Error (MASE) are examples of scale-free metrics. The NRMSE is calculated by: 
\begin{equation}
NRMSE=  \frac{1}{\sigma_{actual}}(\sqrt{\frac{\sum (predicted_i - actual_i)^{2}}{N}})  
\end{equation}
In this context, N represents the total number of samples, $\sigma_{actual}$ denotes the standard deviation of the actual $RSRQ/SNR$ values, and $predicted_i$ and $actual_i$ represent the predicted and actual values for sample $i$.
\par
MASE is a statistical technique that involves comparing the average absolute errors of a model with the average absolute difference of a naïve prediction that replicates the prior value. This measure serves as a valuable tool for evaluating and comparing performance across datasets that include varying sizes. MASE is calculated as:

\begin{equation}
 MASE =   N*\frac{\sum \left| predicted_i - actual_i\right|}{\sum \left| actual_i - actual_{i-1}\right|}
\end{equation}

One example of a scale-dependent metric is the Mean Squared Error (MSE). Although not optimal for cross-dataset comparisons, the MSE may nonetheless provide valuable insights into assessing the overall accuracy of predictions within a particular dataset. It is calculated by:
\begin{equation}
MSE =\frac{ \sum (predicted_i - actual_i)^{2}}{ N}
\end{equation}
In this context, $N$ represents the total number of samples, whereas $predicted_i$ and $actual_i$ denote the predetermined and observed values for sample$ i$, respectively.
\par
The R-squared $(R^{2})$ score quantifies the extent to which the model's predictions account for the variability seen in the actual $RSRQ/SNR$ values. Although not devoid of scale, it offers insight into the degree to which the model aligns with the observed data. The use of NRMSE and MASE enables the evaluation of the model's efficacy across diverse datasets characterized by varied magnitudes of RSRQ and SNR values. MSE may provide valuable insights into the overall accuracy of predictions within our particular dataset.  The $R^{2}$ score is a measure of how well the model captures the fundamental patterns in the data. Evaluating performance across different datasets is essential, and NRMSE and MASE play a vital role in this aspect. While the main emphasis is on comprehending the overall accuracy of predictions within the dataset, the MSE may also be a beneficial tool in conjunction with $R^2$.

\subsection{Evaluation Of RNN-LSTM model }
The graph in Fig. 7 illustrates the convergence of three architectures, namely Convolutional Neural Network (CNN), Recurrent Neural Network (RNN), and RNN-LSTM, in forecasting RSRQ values for a given User located distant from the BS (far user). The models are trained using normalized RSRQ data with a sequence length of 10. Their performance is then compared using RMSE metrics. The objective is to choose the model that produces the most optimal convergence by evaluating their RMSE on a time series dataset for a specified UE. Based on the training RMSE plots, it can be seen that RNN-LSTM has the ability to achieve superior convergence compared to CNN and RNN models. A lower RMSE implies superior model performance since it suggests that the model's predictions are, on average, more accurate and closer to the actual values. This indicates that the RNN-LSTM architecture is more appropriate for capturing the sequential characteristics of RSRQ data and acquiring knowledge of temporal relationships in order to make precise predictions. 
\begin{figure}[htbp]
    \centering
    \includegraphics[width=0.5\textwidth]{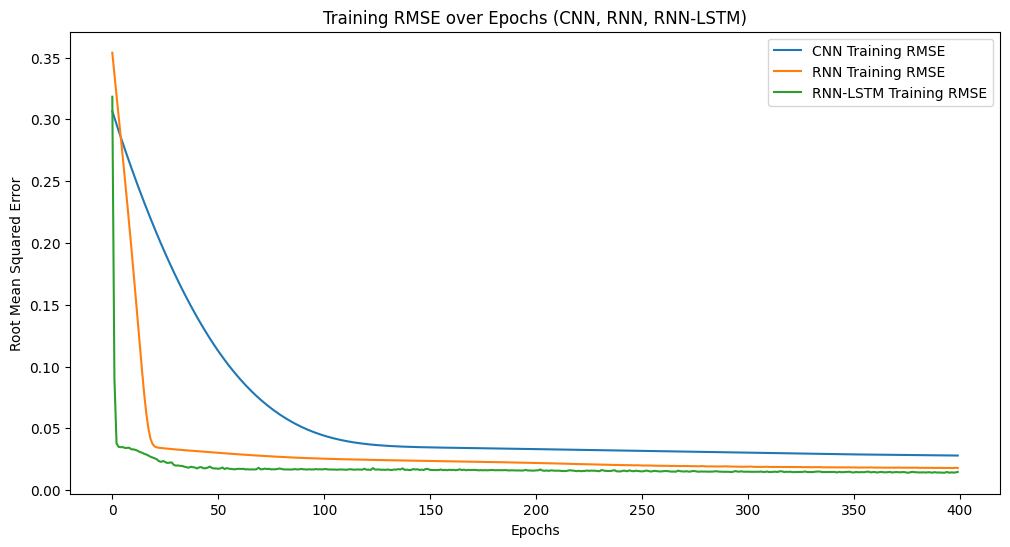} 
    \caption{Training Convergence of three architectures: CNN, RNN and RNN-LSTM}
\end{figure}

This graph in Fig. 8 displays the Normalized Root Mean Squared Error (NRMSE) for predicting RSRQ on two User equipment (UE2 and UE1) located at different distances from the Base Station (BS). The RNN-LSTM model yielded projected RSRQ values that were more closely aligned with the actual values for nearby users. This would result in a decreased NRMSE for the nearby user and an increased NRMSE for the distant user in comparison to the nearby user. UE that is in close proximity to the providing cell tower usually has a greater RSRQ, indicated by a higher number.  UE that is situated at a greater distance from the providing cell tower often encounters a weaker RSRQ, which is indicated by a lower value and eventually more fluctuating NRMSE graph. 

\begin{figure}[htbp]
    \centering
    \includegraphics[width=0.5\textwidth]{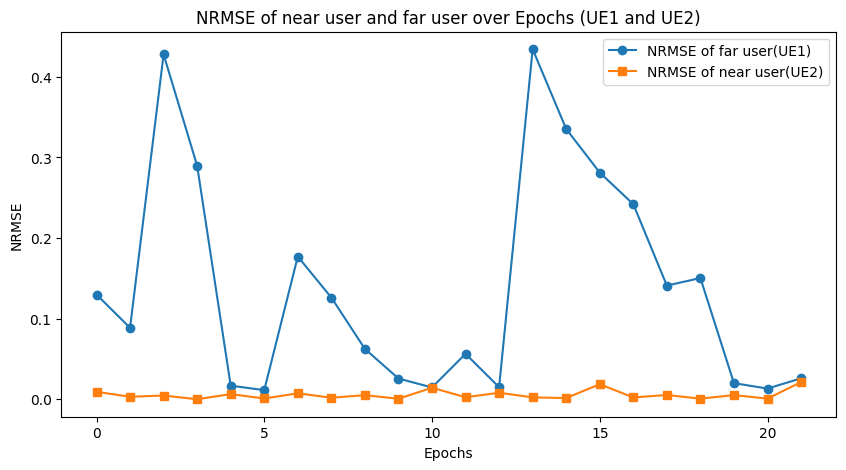} 
    \caption{NRMSE of near user and far user over Epochs (UE2 and UE1)}
\end{figure}

The graph in Fig. 9 illustrates the MAE performance of two situations in forecasting channel information using an RNN-LSTM model. The blue curve indicates the MAE measure when the model is trained on channel data that does not contain partially decoded information, whereas the orange curve reflects the MAE when the model is trained using partially decoded data. It is noteworthy that both curves have variations instead of a completely smooth drop, which is a typical occurrence during the training of LSTM models. The variations occur due to the random characteristics of gradient descent, the optimization process used to train the model. Although there may be some variations, it is preferable to have a continuous decreasing trend in the MAE curve across epochs. Through a comparison of the curves, we can assess the influence of including partly decoded data on the model's convergence and prediction accuracy. If the orange curve (representing partly decoded data) exhibits a more pronounced and persistent reduction in MAE in comparison to the blue curve, it indicates that integrating additional information aids in the model's more efficient learning. In contrast, if the orange curve exhibits comparable or more pronounced variations, it suggests that partly decoded data has little impact on the model's convergence or may potentially contribute unwanted random signals.

\begin{figure}[htbp]
    \centering
    \includegraphics[width=0.5\textwidth]{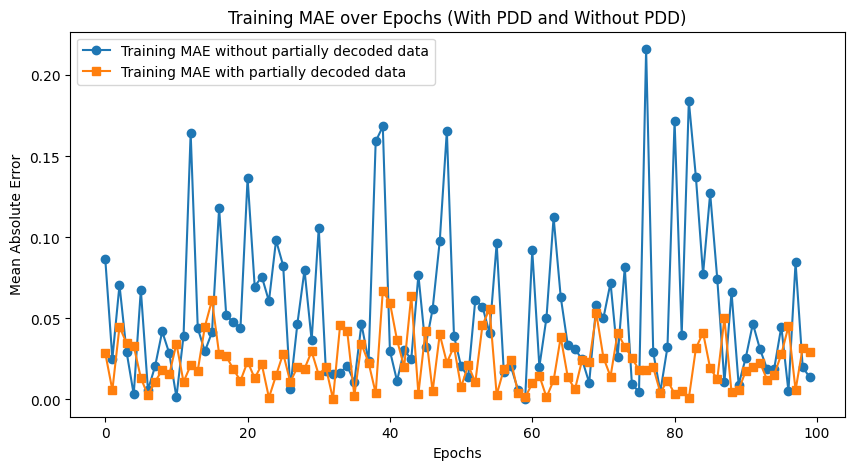} 
    \caption{Training MAE with PDD and without PDD}
\end{figure}

R-squared $(R^{2})$ is a statistical measure used to assess the degree to which a regression model accurately represents a dataset. The coefficient of determination quantifies the amount of variability in the predicted values of the dependent variable that can be attributed to the independent variable(s) included in the model. The $R^{2}$ values vary between $0$ and $1$. A value of $0$ shows no connection between the predicted and actual values, while a value of $1$ indicates a perfect fit, where the model accurately predicts the dependent variable. By including PDD information, the model achieves a more accurate alignment between the predicted and actual channel information, demonstrating its enhanced effectiveness. The inclusion of PDD in the model is anticipated to enhance its ability to account for the effects of signal delays on CSI, hence resulting in more precise predictions. By processing the raw signal to extract PDD characteristics, the model gains more comprehensive knowledge about the behavior of the signal. This eventually leads to a deeper understanding of the link between the features and the target state of CSI. By visually comparing the graphs in Fig. 10, we can verify if the model trained with PDD regularly achieves a better $R^{2}$ compared to the model without PDD. If the errors in partially decoded data are excessive or incomprehensible, they may introduce noise rather than useful information, resulting in a paradoxical scenario. This might result in a decreased $R^{2}$ in comparison to only using CSI. In the NOMA network we are studying, a user located far away may experience an abrupt change in the R-squared value during training. This might be due to difficulties in accurately recording the signal behavior for remote users. If the learning algorithm of the model can detect patterns that are unique to distant user signals, it may encounter a dramatic increase in R-squared. This indicates the enhanced capability of the model to address the difficulties related to predicting signals from distant users.

\begin{figure}[htbp]
    \centering
    \includegraphics[width=0.5\textwidth]{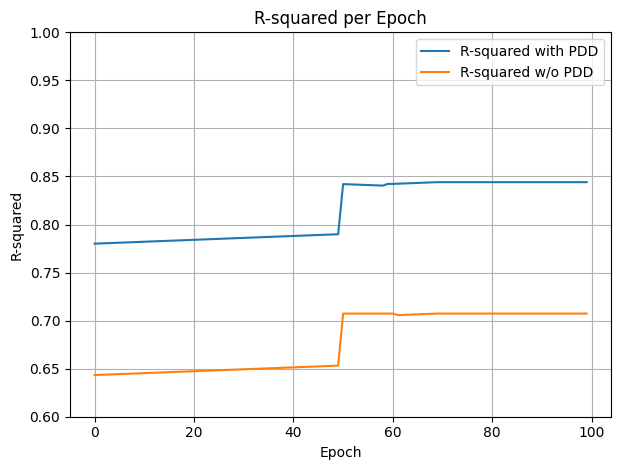} 
    \caption{R-squared comparison with PDD and without PDD}
\end{figure}

The R-squared score curve in Fig. 11 demonstrates a distinct ranking in the effectiveness of different models in predicting RSRQ results. The RNN-LSTM model regularly attains the greatest R-squared scores throughout the training phase, showing its better capacity to capture the underlying connections between the input data and the target variable.  The training data may include a fresh data point or a group of points that have a noticeable and well-defined connection to RSRQ at that particular time period. The CNN uses 1D convolution with 32 filters and utilizes the Rectified Linear Unit (ReLU) activation function to extract features. The RNN model uses a single LSTM layer with 20 units and a ReLU activation function to capture the temporal dependencies present in the data effectively. The RNN-LSTM model consists of two LSTM layers, with the first layer having 50 units and the second layer having 20 units. The ReLU activation function is used to facilitate the learning of intricate associations. Nevertheless, the constraints of the CNN design result in negative R-squared values. This suggests a worse match compared to the reference point.
\begin{figure}[htbp]
    \centering
    \includegraphics[width=0.5\textwidth]{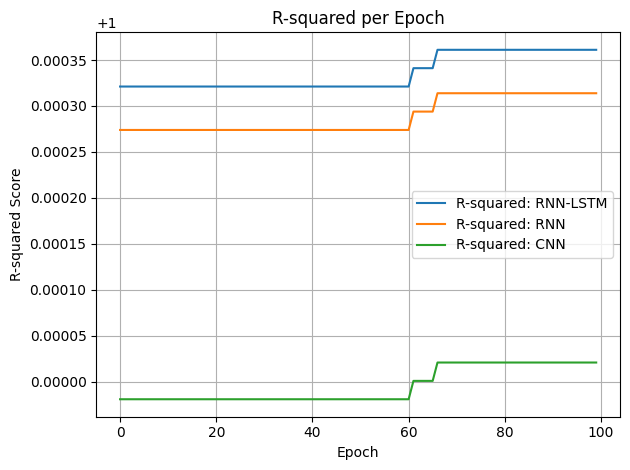} 
    \caption{R-squared Score (without PDD): CNN, RNN, RNN-LSTM}
\end{figure}
The findings and discussion center on the influence of using PDD to evaluate the status of the channel. The RNN-LSTM model makes predictions on the Signal-to-interference-plus-Noise Ratio (SINR), which is an important measure of channel quality. The quantity of data frames defines the quantity of training data used for prediction. Fig. 12 depicts the training patterns of models that use CSI exclusively and those that include PDD as input. The CSI-exclusive model shows a curve that falls below 0 in some epochs. This suggests that the model may not be well-suited for some data points, indicating that it has difficulty capturing the intricacies of the channel without more information. On the other hand, the model that includes partially decoded data regularly produces a curve that is higher than 1, indicating a more pronounced positive relationship with the goal values. Fig. 13 of the testing comparison between SINR values illustrates the superiority of taking PDD as CSI, which results in a better SINR output.  This emphasizes the advantage of PDD in improving the model's comprehension of channel fluctuations. Comparable trends are noted while comparing testing data sets.

\begin{figure}[htbp]
    \centering
    \includegraphics[width=0.5\textwidth]{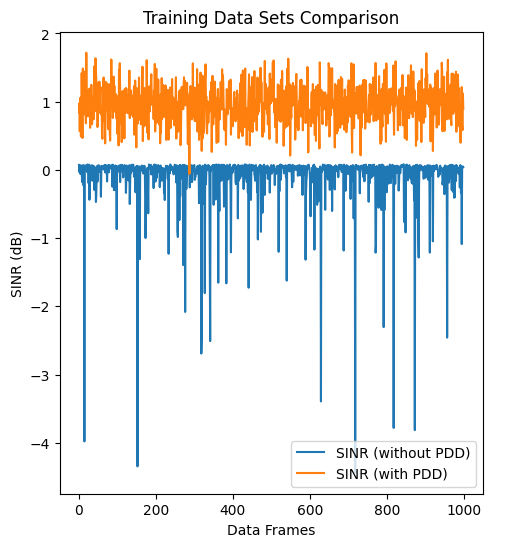} 
    \caption{Training SINR Comparison over Data Frames}
\end{figure}

\begin{figure}[htbp]
    \centering
    \includegraphics[width=0.5\textwidth]{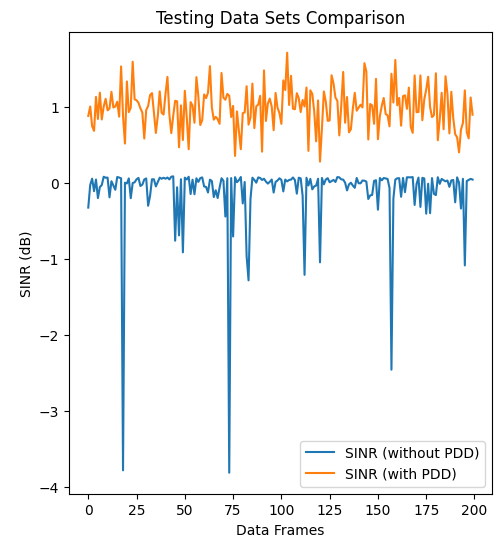} 
    \caption{Testing SINR Comparison over Data Frames}
\end{figure}
We employ scale-free and scale-dependent measures, such as RMSE, MAE, and MSE scores in TABLE IV, to assess the behavior of various models on a dataset. Table IV lists the performance metrics for several ML models, such as CNN, RNN, and RNN-LSTM. Table IV presents the accuracy superiority of the proposed RNN-LSTM model over the two remaining models, namely CNN and RNN, based on different datasets from \cite{15}. The complete code related to simulation can be found in \cite{16}.
\begin{table}
\caption{Performance metrics of different ML models}
\begin{center}
\scriptsize
\begin{tabular}{||p{0.7cm}|p{0.5cm}|p{0.6cm}|p{0.5cm}|p{0.6cm}|p{0.5cm}|p{0.6cm}|p{0.5cm}||}
\hline
Metrics &  Models & Amazon Prime  &  & Download & & Netflix &\\ 
\hline
  & & Driving & Static & Driving & Static & Driving & Static \\ 
 \hline
 RMSE  & CNN RNN  RNN-LSTM &  0.0798 0.0722 0.0608 & 0.0364 0.0323 0.0119 & 0.1092 0.0845 0.0737 & 0.1889 0.1478 0.1322  & 0.1116 0.0764 0.0685 &  0.1056 0.0837 0.0712\\ 
 \hline
  MAE  & CNN RNN  RNN-LSTM & 0.0695 0.0585 0.0386  & 0.0069 0.0065 0.0021 & 0.0949 0.0543 0.0502 & 0.1543 0.1014 0.0922 & 0.0816 0.0489 0.0343 & 0.0853 0.0541 0.0429 \\ 
 \hline
  MSE  & CNN RNN  RNN-LSTM & 0.0075 0.0054 0.0034  &  0.0013 0.0009 8.3417e-01 & 0.0136 0.0075 0.0063 & 0.0346 0.0218 0.0173 & 0.0117 0.0068 0.0046 & 0.0095 0.0062 0.0047 \\ 
 \hline
 \end{tabular}
\end{center} 
 
\end{table}
 \subsection{Application of the proposed technique to HLF/RLF}
This research aims to examine the influence of handover decision-making, which is based on changing CSI, on the frequencies of handover failures. We use five handover profiles obtained from \cite{14}, each of which may reflect different network setups defined by specific values for Time to Trigger (TTT), $A3$ Offset, and $L3$ Filter $K$.  Next, we model handover situations for UE moving at different velocities. Each profile is evaluated using two CSI configurations: 1) using only RSRQ and 2) employing both RSRQ and PDD. The generated graph effectively illustrates how the selection of CSI affects handover performance. This is done in Fig. 14 by comparing the handover failure rates for these two CSI setups at various UE speeds and handover profiles. This analysis will provide valuable insights into optimizing handover strategies based on the available CSI and network conditions. 
\begin{figure}[htbp]
    \centering
    \includegraphics[width=0.40\textwidth]{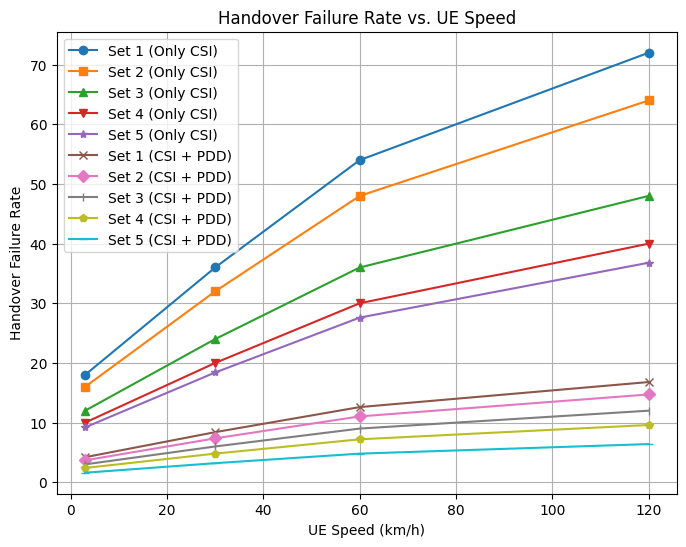} 
    \caption{Handover failure rates over UE speeds (km/hour)}
    \hfill
\end{figure}
We examine how handover decision-making affects ping pong rates (handover frequency) for UEs moving at different speeds. Five handover profiles from \cite{14} reflect various network settings. We evaluate two CSI setups for each profile. We have given a complete graph in Fig. 15 that shows how CSI and network design (as specified by handover profiles) affect handover frequency by comparing ping-pong rates across various UE speeds, handover profiles, and CSI configurations. PDD provides more real-time channel quality information than signal strength (RSRQ), which improves handover choices and reduces ping pong. Handover failures arise from inaccurate forecasts of forthcoming channel conditions. Unsuccessful transfers might result in a consistent pattern in the curves that represent the rate of failure in transferring responsibilities. The ping pong rate takes into account both successful and unsuccessful handovers, which might result in less predictable patterns. The setup of the network and the information about the PDD may have a significant influence on both the success and failure rates, therefore affecting the distribution of ping-pong rates.
\begin{figure}[htbp]
    \centering
    \includegraphics[width=0.40\textwidth]{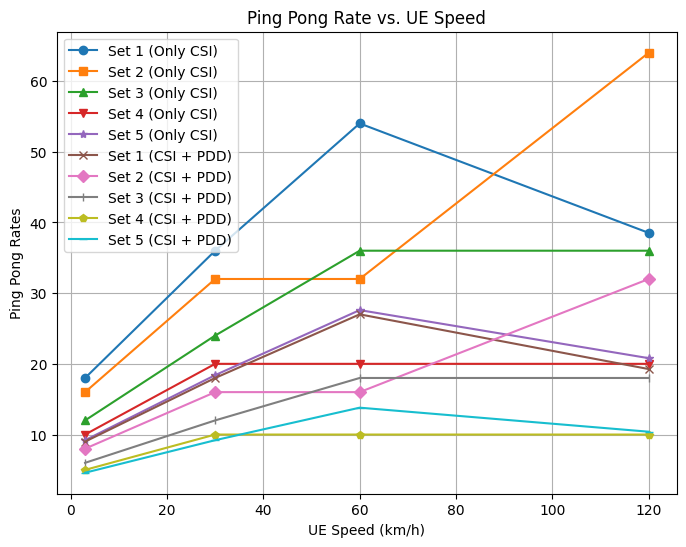} 
    \caption{Ping-pong rates over UE speeds (km/hour)}
   \end{figure}
The analysis of false alarm rates reveals a U-shaped curve when plotted against UE speeds. At very low speeds, even minor fluctuations in the channel, which might appear significant, can trigger unnecessary handover attempts (false alarms) due to limitations in predicting future channel conditions based solely on Received Signal Strength Received Quality (RSRQ). However, as UE speeds increase dramatically, the channel environment changes rapidly, making frequent handovers inevitable to maintain a good connection. This contrasting behavior at both ends of the speed spectrum leads to a peak in the number of false alarms at moderate speeds. Here, the challenge lies in accurately distinguishing between temporary fluctuations and actual signal degradations, making handover decisions more critical. The graph in Fig. 16 incorporates two curves representing the impact of CSI used for handover decisions, one only relying on RSRQ and another one considering both RSRQ and PDD. The data suggests that utilizing both CSI and PDD leads to generally lower false alarm rates compared to relying solely on RSRQ across most UE speeds.
\begin{figure}[htbp]
    \centering
    \includegraphics[width=0.5\textwidth]{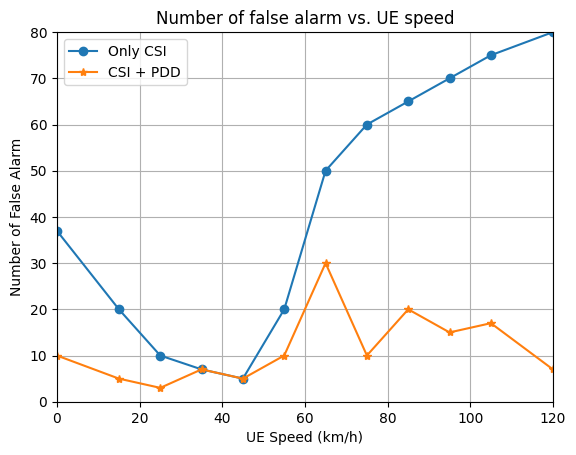} 
    \caption{Number of false alarm vs. UE speeds (km/hour)}
   \end{figure}

\section{Conclusion and Future works}
A novel method for NOMA channel prediction was introduced in this study, which makes use of a larger variety of CSI parameters, such as RSRP, RSRQ, CQI, SNR, and, most importantly, PDD. System-level simulations showed that low CSI has a major effect on radio link and handover failures. Our results demonstrate the benefits of PDD as a supplementary CSI metric.  By doing away with the requirement for specialized pilot signals, PDD lowers signaling overhead while providing a more realistic depiction of channel dynamics than conventional techniques. Additionally, this method is adaptable to shifting user traffic patterns. In comparison to a deterministic model, our suggested approach performed better in terms of MSE and BER. This is the first study to investigate the application of PDD in NOMA networks to forecast handover and radio link failures. Realizing that ML can be used for channel estimation because it doesn't require as many strict assumptions. In addition, we used a transfer learning strategy to get around dataset size constraints. Furthermore, we exhibited the competitive performance of our model in comparison to earlier research and investigated the use of scale-free and scale-dependent metrics. By using PDD as a useful source of channel information for forecasting channel behavior in NOMA networks, this study sets itself apart. In subsequent deployments, this strategy could enhance network performance and handover decision-making. Future objectives for the study include examining the effects of PDD on network performance under different traffic scenarios and delving deeper into the incorporation of ML algorithms for channel estimation.

\section*{Acknowledgment}
The authors would like to express their sincere gratitude to Kaushal Shelke from the Department of Computer Science and Engineering, Indian Institute of Information Technology Bhagalpur, and Ayush Agarwal from the Department of Computer Science and Engineering, Indian Institute of Technology Guwahati, for their valuable contributions to this paper. Their insightful and dedicated efforts have significantly enhanced the quality of this research.

\end{document}